\documentclass{ws-procs975x65}
\usepackage{graphicx}

\begin{document}
  
\title{\uppercase{The EBEX Cryostat and Supporting Electronics}}

\author{ILAN SAGIV, ASAD M. ABOOBAKER, CHAOYUN BAO, SHAUL HANANY, TERRY JONES, JEFFREY KLEIN, MICHAEL MILLIGAN, DANIEL E. POLSGROVE, KATE RAACH and KYLE ZILIC}
\address{University of Minnesota School of Physics and Astronomy\\Minneapolis, MN 55455, USA}

\author{ANDREI KOROTKOV, GREGORY S. TUCKER and YURY VINOKUROV}
\address{Brown University\\Providence, RI 02912, USA}

\author{TOMOTAKE MATSUMURA}
\address{California Institute of Technology\\Pasadena, CA 91125, USA}

\author{PETER ADE, WILL GRAINGER and ENZO PASCALE}
\address{Cardiff University\\Cardiff, CF24 3AA,United Kingdom}
                                                 
\author{DANIEL CHAPMAN, JOY DIDIER, SETH HILLBRAND, BRITT REICHBORN-KJENNERUD, MICHELE LIMON and AMBER MILLER}
\address{Columbia University\\New York, NY 10027, USA}

\author{ANDREW JAFFE}
\address{Imperial College\\London, SW7 2AZ, England,United Kingdom}

\author{AMIT YADAV and MATIAS ZALDARRIAGA}
\address{Institute for Advanced Study\\Princeton, NJ 08540, USA}

\author{NICOLAS PONTHIEU}
\address{Institut d'Astrophysique Spatiale, Universite Paris-Sud\\Orsay, 91405, France}

\author{MATTHIEU TRISTRAM}
\address{Laboratoire de l'Acc\'el\'erateur Lin\'eaire, Universit\'e Paris Sud, CNRS, Orsay, France}

\author{JULIAN BORRILL, CHRISTOPHER CANTALUPO and TED KISNER}
\address{Lawrence Berkeley National Laboratory\\Berkeley, CA 94720, USA}

\author{FRAN\c COIS AUBIN, MATT DOBBS and KEVIN MACDERMID}
\address{McGill University, Montr\'eal\\Quebec, H3A2T8, Canada}

\author{GENE HILTON, JOHANNES HUBMAYR, KENT IRWIN and CARL REINTSEMA}
\address{National Institute of Standards and Technology\\Boulder, CO, 80305, USA}

\author{CARLO BACCIGALUPI and SAM LEACH}
\address{Scuola Internazionale Superiore di Studi Avanzati\\Trieste 34014, Italy}

\author{BRADLEY JOHNSON, ADRIAN LEE and HUAN TRAN}
\address{University of California, Berkeley\\Berkeley, CA 94720, USA}

\author{LORNE LEVINSON}
\address{Weizmann Institute of Science\\Rehovot 76100, Israel}

\begin{abstract}
We describe the cryostat and supporting electronics for the EBEX experiment. EBEX is a balloon-borne polarimeter designed to measure the B-mode polarization of the cosmic microwave background radiation. The instrument includes a $1.5$ meter Gregorian-type telescope and 1432 bolometric transition edge sensor detectors operating at 0.3 K. Electronics for monitoring temperatures and controlling cryostat refrigerators is read out over CANbus. A timing system ensures the data from all subsystems is accurately synchronized. EBEX completed an engineering test flight in June 2009 during which the cryogenics and supporting electronics performed according to predictions. The temperatures of the cryostat were stable, and an analysis of a subset of the data finds no scan synchronous signal in the cryostat temperatures. Preparations are underway for an Antarctic flight.
\end{abstract}


\bodymatter

\section{Introduction}\label{sec:Science}
EBEX (E and B EXperiment) is a balloon-borne experiment equipped with 1432 bolometric transition edge sensors (TES). It is designed to measure the temperature and polarization of the cosmic microwave background (CMB) radiation. EBEX has several scientific goals. The first is to search for the B-mode polarization of the CMB which is a signature of inflationary gravitational waves [\citen{kamionkowski97b},\citen{seljak97}]. The second goal is to characterize Galactic polarized dust emission and to determine its angular power spectra in both E and B-mode polarizations at the micro-Kelvin level. Characterization of dust foregrounds is a key element in extracting the B-mode signal [\citen{brandt94},\citen{baccigalupi03}]. The third goal is to measure the predicted but yet undetected lensing of the CMB. Gravitational lensing of the CMB distorts the anisotropy pattern on the sky, modifies its power spectrum [\citen{seljak97a}], and converts E-mode to B-mode polarization [\citen{zaldarriaga98}]. 

Previous EBEX related publications describe the experiment's science goals [\citen{Oxley04}], design [\citen{granger08}] and detectors and readout [\citen{hubmayr08}]. Here we will focus on the cryostat and its housekeeping readout.
\begin{figure}
\label{fig:EBEX_payload}
\begin{center}	
  	\parbox{2.1in}{\epsfig{file=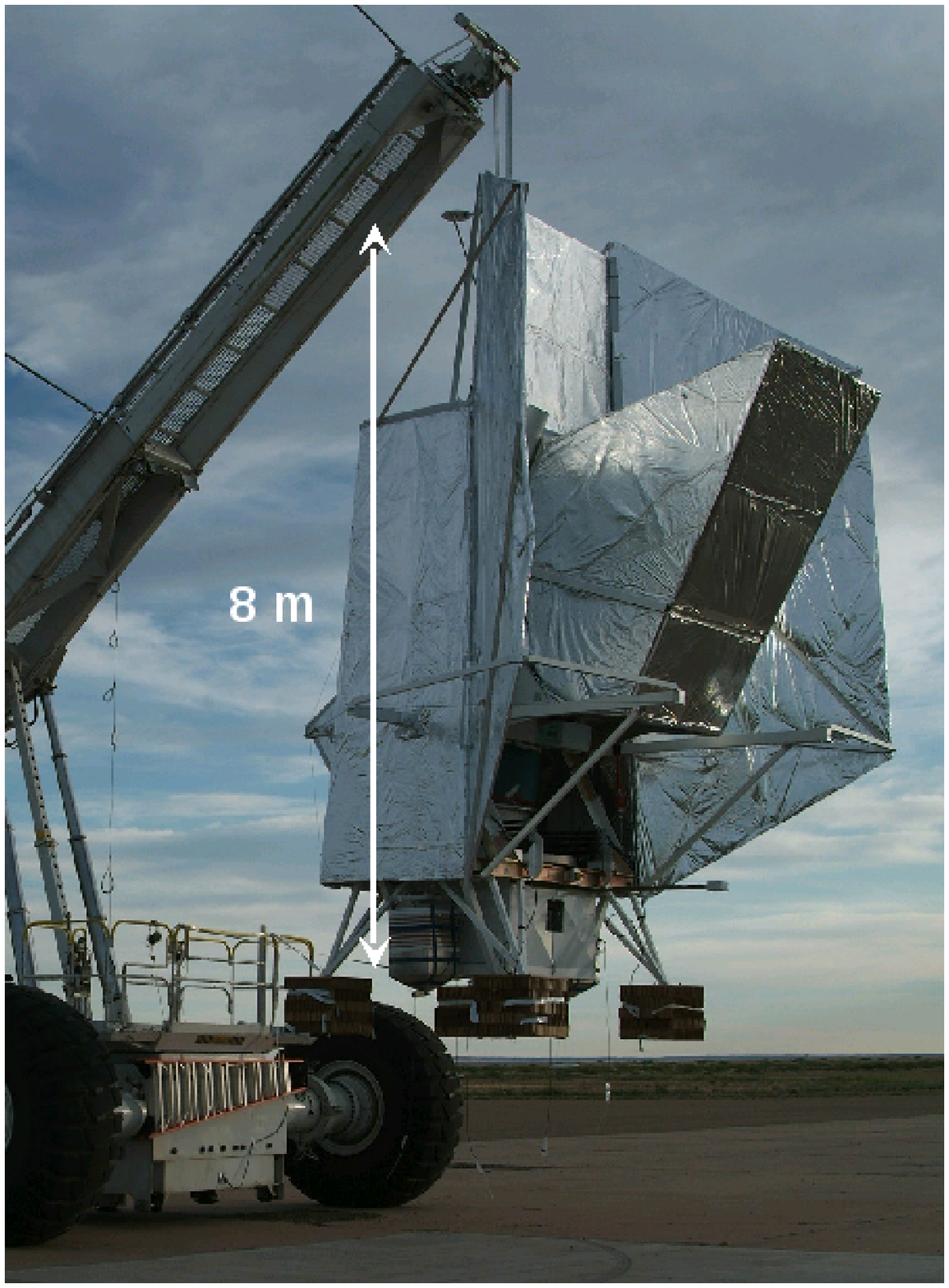,width=2in }}
  	\hspace*{4pt}
  	\parbox{2.1in}{\epsfig{file=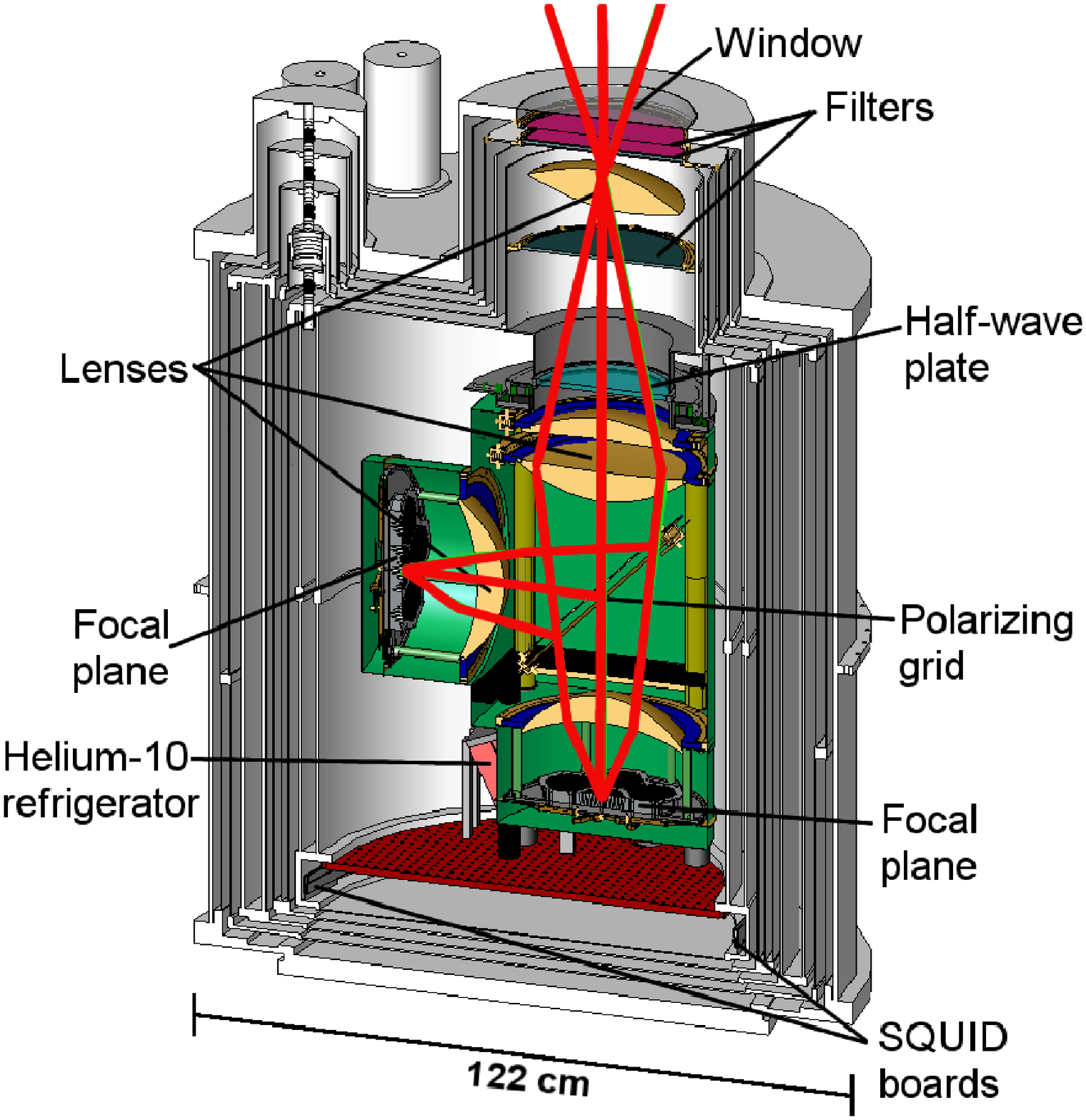,width=3in}}
  	\caption{Left: The EBEX payload before launch. Right: A cross section of the EBEX cryostat.}
\end{center}
\end{figure}

\section{Overview\label{sec:Experimental-Design}}
EBEX will be flown over Antarctica where it will scan $\sim$1\% of the sky with a resolution of 8$'$. Fig. 1 shows the EBEX payload and a cross section view of the cryostat. The EBEX optical system consists of a 1.5~m aperture primary mirror and a 1~m elliptical secondary, that together make a Dragone-type telescope (hidden behind baffles in the left panel of Fig. 1). Light focused into the cryostat passes several optical elements before reaching one of two focal planes (Fig. 1 - right panel).

\section{The Cryostat}
\subsection{Overall Design}

The cryostat [\citen{precCryo}] which holds the cold optics and detectors has a diameter of 122~cm and is 150~cm tall, with an optics snout extending an additional 30~cm (Fig. 1, right). The toroidal liquid helium tank has a capacity of 67~liters that is surrounded by a liquid nitrogen tank containing 120~liters. The cryostat also has two intermediate vapor cooled shields operating at $\sim$30~K and $\sim$185~K. The cryogens are kept at atmospheric pressure by an absolute pressure regulator valve and are designed to last 21 days.

\begin{figure}
\begin{center}
\psfig{file=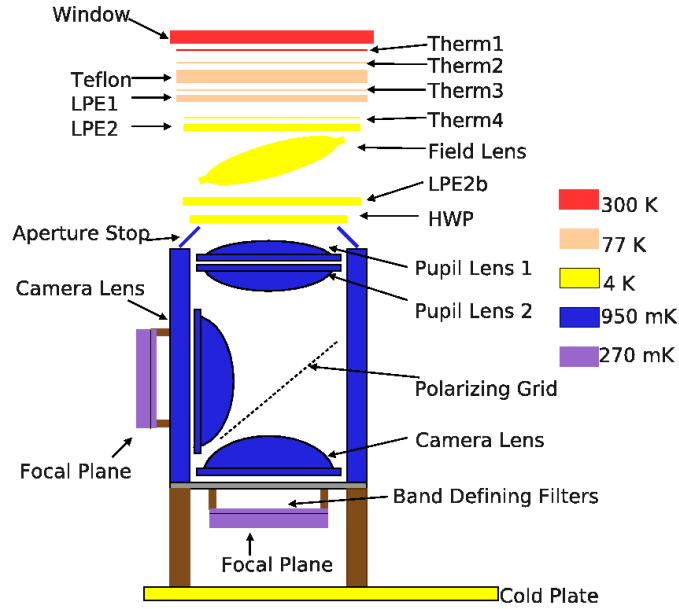,width=3.5in}
\caption{EBEX cold optics color coded by temperature}
\label{fig:optics_color}
\end{center}
\end{figure}

A half-inch thick ultra-high molecular weight polyethylene (UHMWPE) window is mounted at the entrance to the evacuated cryostat. In order to reduce optical loading from this window during the Antarctic flight, it will be removed from the optical path by a motorized mechanism, leaving a thin UHMWPE window. This will happen once the experiment reaches float altitude, where the pressure differential is a few Torr. A series of filters [\citen{cardiff}] shown in Fig. \ref{fig:optics_color} are designed to prevent high-frequency thermal radiation from reaching the cold optics and the detectors. Three types of filters are used: `thermal' filters with low emissivity and cut-off frequencies of 60~THz, 30~THz, 15~THz and 12~THz (Therm1, Therm2, Therm3 and Therm4, respectively), `Low-Pass Edge' (LPE) filters with cut-off frequencies at 750 GHz, 600 GHz and 540 GHz (LPE1, LPE2 and LPE2b, respectively), and a half inch thick Teflon low-pass filter which has low emissivity in the EBEX bands. The Teflon filter has absorption greater than 90\% at frequencies above 10~THz.

An achromatic half-wave plate (HWP)~[\citen{hanany05},\citen{matsumura08}] is mounted on the sky side of a 22~cm diameter aperture stop (Fig. \ref{fig:optics_color}). The HWP is continuously rotated at 2 Hz on a superconducting magnetic bearing (SMB)[\citen{Hanany03}] by a Kevlar belt that is driven by a DC motor mounted outside the cryostat. The angular orientation of the HWP is encoded with a slotted ring illuminated by a laser. We have demonstrated reconstruction of the HWP angular orientation to better than $0.1^\circ$ accuracy [\citen{jk10}].

A 43.5~cm diameter polarizing wire grid is mounted at $45^\circ$ relative to the incident radiation, splitting the beam to the two focal planes. Distributed along the optical path are four lenses per focal plane, all made of a single block of UHMWPE, which provide a flat and telecentric focal plane suitable for the TES arrays. The lenses are coated with a broad-band anti-reflection coating. Light is coupled to the bolometers by an array of smooth-walled conical horns. Band defining filters are mounted before each wafer on the sky side of the horn array. Each focal plane has three frequency bands centered at 150~GHz, 250~GHz and 410~GHz with a total of 768, 384, and 280 detectors in each band respectively. The design parameters of the bolometers are optimized for the lower optical loading at balloon altitudes. The bolometers are read out using a novel low power digital frequency multiplexing scheme and series array SQUID amplifiers [\citen{huber01_squid},\citen{dobbs08}]. Each SQUID reads out 12 bolometers, thus limiting the total number of wires required between the cryogenic stages to 1110. These wires, that control the SQUIDs and bring the bolometer bias and output signals in and out of the cryostat, are made of manganin weaved into Nomex ribbons [\citen{tekdata}]. The load on the liquid He stage from these ribbons is 30~mW.

The lenses, grid and supporting structure are cooled with a two-stage $^4$He/$^4$He adsorption refrigerator which maintains a temperature of $\sim$950~mK with a load of 200~$\mu$W for $\sim$4~days [\citen{chase}]. The bolometers and focal planes are cooled using a three-stage $^4$He/$^3$He/$^3$He adsorption refrigerator which maintains a temperature of $\sim$270~mK with a load of 6~$\mu$W for $\sim$5~days. Fig. \ref{fig:optics_color} shows the cold optics color coded by temperature.

\subsection{RF Mitigation}
\begin{figure}
[top]
\begin{center}
\epsfig{file=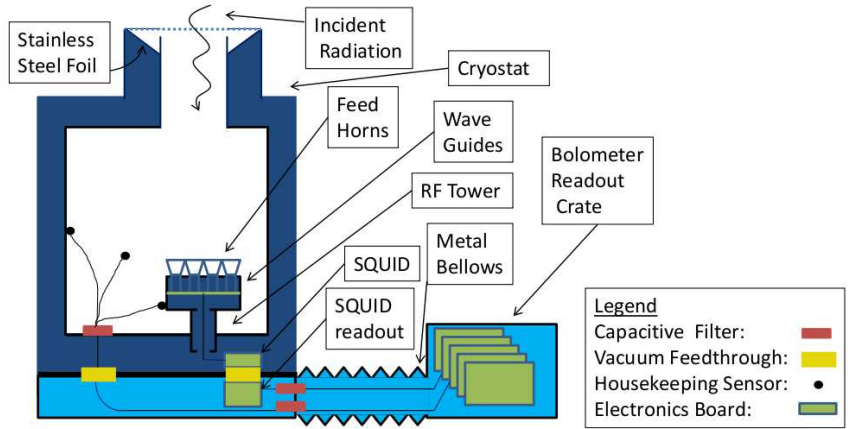,width=4.5in}
\caption{Schematic of the EBEX RF mitigation strategy. Blue (Cyan) areas are RF shielded inside (outside) the cryostat.}
\label{fig:rf}
\end{center}
\end{figure}

The design for RF mitigation is shown schematically in Fig. \ref{fig:rf}. The inner volume of the cryostat is not considered RF clean due to the large window. Indium seals and RF gaskets [\citen{gasket}] are used to keep all other parts of the cryostat RF clean. To prevent RFI from entering between the cryostat layers while not compromising the cryogenic performance, 25.4~$\mu$m thick stainless steel sheets are wrapped around the optics snout. Wires that penetrate the RF tight area are filtered using capacitive low-pass filters [\citen{capfilt},\citen{hwpfilt}]. A high-pass wave guide with a cut-off frequency at 230~GHz is installed on the cold plate (Fig. \ref{fig:optics_color}) to allow pumping out of the inner volume. The conical horns, coupling the light to the focal-plane, place the detectors inside a Faraday cage where they are shielded from electromagnetic interference which may enter the cryostat window. A thermally isolating `RF tower' is created from Vespel wrapped with 5~$\mu$m thick superconducting niobium foil. Detector wiring is routed through this tower to keep the wiring in the RF clean area. 

The detector wiring from the cryostat to the readout crates is wrapped in bronze wool and is routed through a flexible metal hose. 

\section{Cryostat Housekeeping Electronics}
We use four types of boards to monitor the temperatures and to control the adsorption refrigerators in the cryostat. The boards are interconnected and controlled over a Controller Area Network (CANbus). CANbus is a high-reliability, low speed, 2-wire serial bus for industrial control. The CANbus is connected to the two flight computers via CANbus-to-USB interfaces [\citen{usb2can}] using the CANopen communication protocol. Each board is managed by an Embedded Local Monitoring Board (ELMB) - a general purpose I/O microprocessor plug-on daughter board that has a CAN interface. The ELMB was developed for the ATLAS experiment at CERN~[\citen{atlas08}] and is radiation tolerant. 

\subsection{General Housekeeping board}
A general housekeeping board is used as an interface to the ELMB's 24 digital I/O channels and 64, 16-bit differential analog channels. Channels can be used to read analog voltages or temperature sensors. Digital I/O channels are used to control the HWP system. During the Antarctic flight these boards will also be used to monitor temperatures and voltages of the solar panel and battery charging system.

\subsection{Silicon Diode Board}
\label{sec:sid}
\begin{figure}
\begin{center}
\epsfig{file=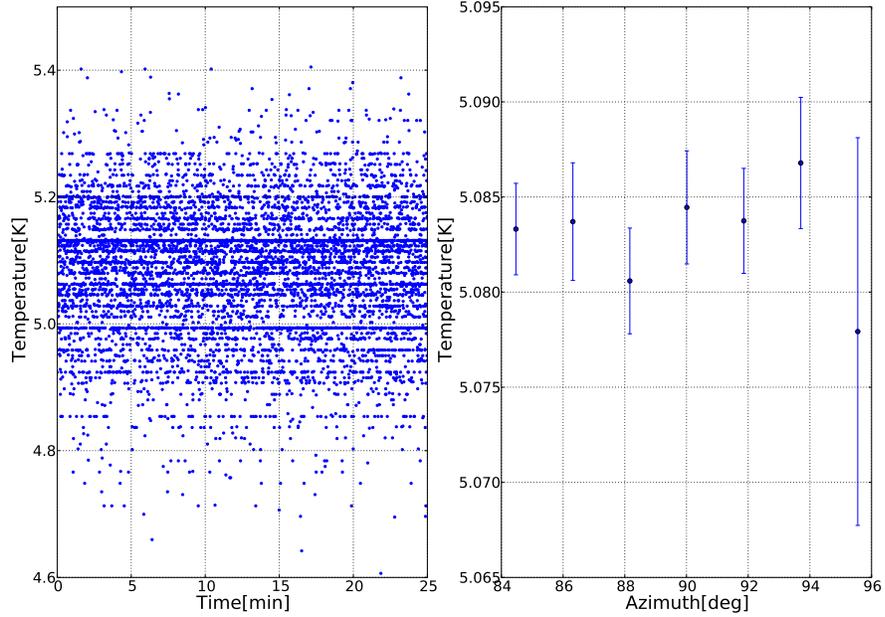,width=5.5in}
\caption{Left: A measurement of the temperature of the liquid helium tank during a sinusoidal azimuth scan with a peak-to-peak amplitude of $9^\circ$ and a period of 25~s. Right: The time ordered data from the left panel binned and plotted vs. azimuth.}
\label{fig:sid}
\end{center}
\end{figure}

To measure temperatures above 1.5~K we use silicon diode temperature sensors~[\citen{sid}]. EBEX has a total of three silicon diode boards; each has 16 channels with a 10~$\mu$A DC excitation current and consumes 0.7~W. The left panel in Fig. \ref{fig:sid} shows a 25~minute segment of temperature data for a typical sensor from the top of the liquid He tank during a scan. The noise of the diode readout is 3~mV~rms which corresponds to 0.1~K at liquid He temperature. The right panel in Fig.~\ref{fig:sid} shows the temperature of the same sensor binned and plotted vs. telescope azimuth. We search for a sinusoidal scan-synchronous temperature modulation that has a period of twice the scan size. We place an upper limit of 9~mK on the amplitude, limited by statistical uncertainty. 

\subsection{Germanium/Ruthenium Readout Board}

\begin{figure}
\begin{center}
\epsfig{file=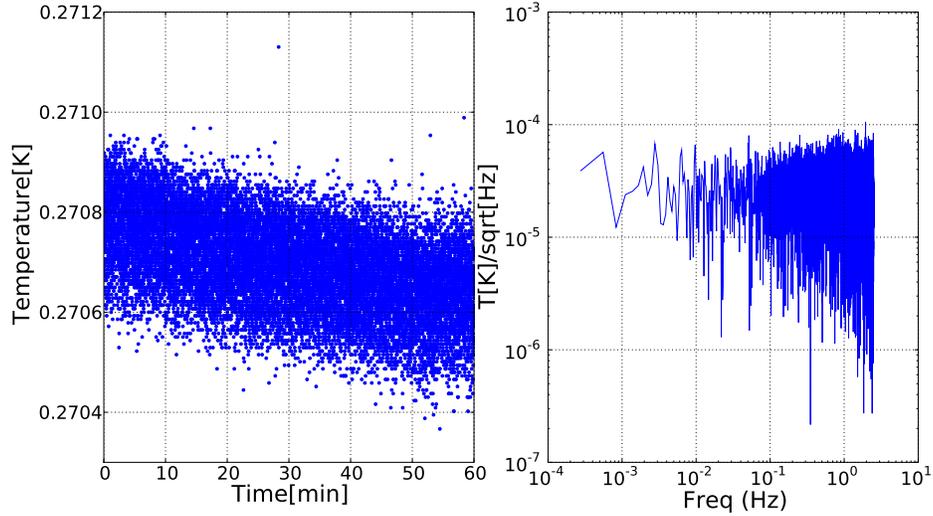,width=5.5in}
\caption{Left: A measurement of the temperature of the 410~GHz wafer during an azimuth scan with a peak to peak amplitude of up to 120$^\circ$ at a scan speed of 0.5$^\circ$/s. Right: Power spectrum of the temperature data after removing an offset and gradient.}
\label{fig:grt}
\end{center}
\end{figure}

For measuring sub-Kelvin temperatures we use Ruthenium dioxide temperature sensors [\citen{rux}]. Each board has seven 4-wire measurement channels plus a temperature sensor measuring the board's temperature. A board consumes 1.6~W. The sensors are excited with a square wave at a frequency of 250~Hz dissipating 60~pW at the sensor when the resistance is $\sim$6~k$\Omega$ ($\sim$280~mK). The voltage sense signal is demodulated, filtered and digitized. In EBEX two boards are used for measuring the temperatures of the adsorption refrigerator's cold stages, the temperatures of the TES wafers and the focal plane structure. The left panel in Fig.~\ref{fig:grt} shows the temperature from a sensor mounted on one of the focal plane wafers during a wide azimuth scan. The noise level is 50~$\mu$K rms. The amplitude of a scan synchronous signal is limited to 7~$\mu$K in the same method described in section~\ref{sec:sid}. The right panel in Fig.~\ref{fig:grt} shows the power spectrum of this sensor showing a flat response down to a 10 milli-Hertz level.

\subsection{Heater Board}
The heater board is used to cycle the adsorption refrigerators. The board has eight channels designated for pump heaters and six channels for gas-gap heat switches. Each heater channel has three states: High (0 - 24 V), Low (0 - 6 V) or Off, the value of which is set by a potentiometer. Each heat switch channel can be preset to an output voltage of up to 5.5 V. The state of each heater or heat switch channel is determined by a solid state relay which is controlled by a programmable logic device (PLD). The PLD is controlled by the ELMB microprocessor which receives the refrigerator state command from the flight computer over the CANbus. The board itself consumes 0.75~W and has an external DC/DC board that supplies power for the heater channels. All output currents are monitored via a sense resistor and are read out over the CANbus. 

\section{Timing}
The bolometer data is recorded at a rate of 190.73 Hz. Pointing sensors are sampled at 5.008~Hz and 100.16~Hz. Errors in matching of the sampling times between the two systems lead to systematic errors. We determined that an accuracy of 1~ms in the relative timing between bolometer and pointing sensors is required to achieve our science goals. The EBEX timing system is designed to provide relative timing that is better than 10~$\mu$s.

\begin{figure}
\begin{center}
\psfig{file=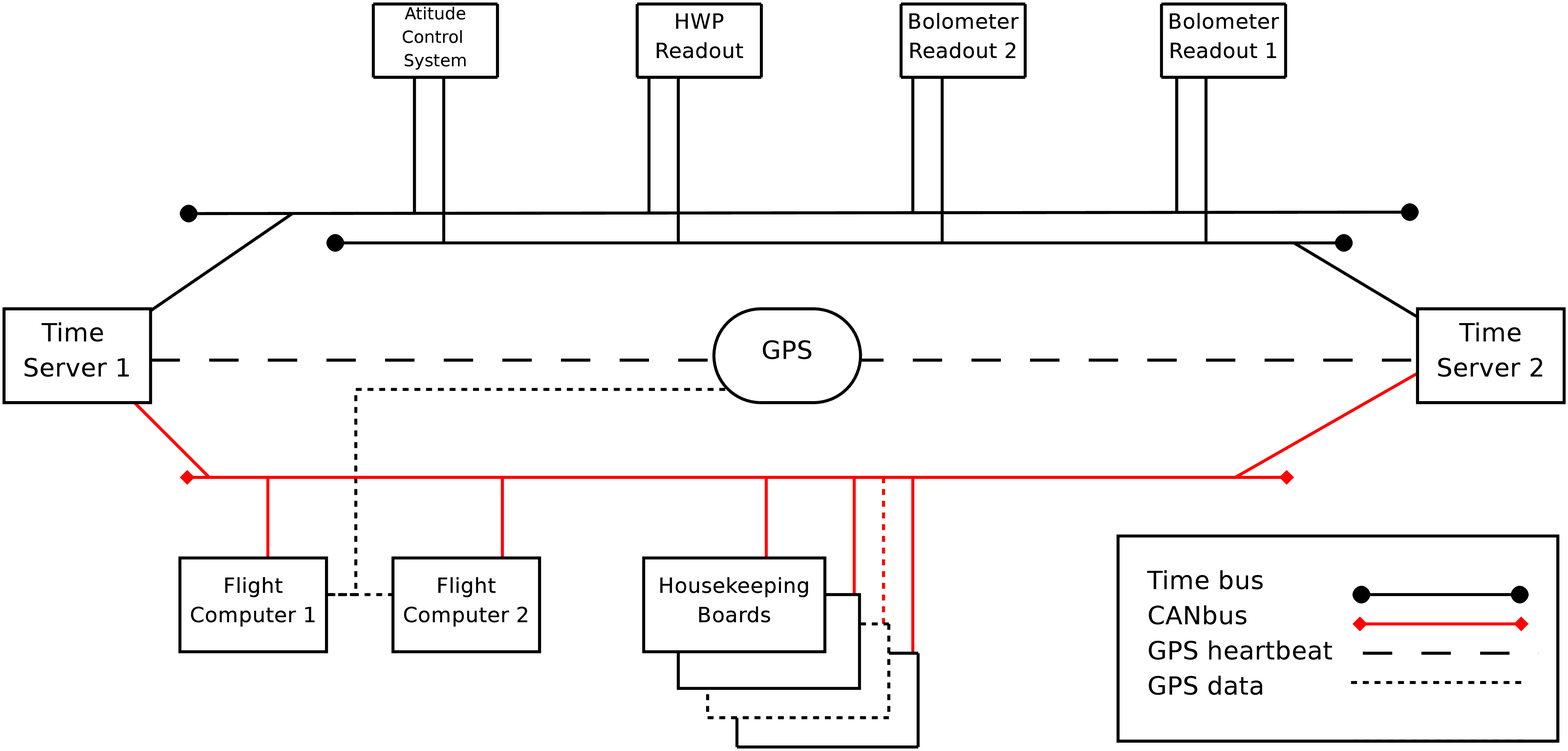,width=5.3in}
\caption{The EBEX time bus. The black line represents the RS-485 time buses.}
\label{fig:timebus}
\end{center}
\end{figure}

The timing system consists of two redundant time servers and the following clients: the bolometer and HWP readout boards; the attitude control system; and each of the two flight computers. The time servers have an oven-controlled oscillator [\citen{vpf}] with a stability of $\pm0.2$~parts per billion. The clients each have oscillators with stability of $\pm25$~parts per million. Each of the time servers communicates to the clients through a separate RS-485 serial link, making two redundant time buses (see Fig. \ref{fig:timebus}). Firmware on the clients sets the conditions for which of the two servers is used.

Each bolometer, attitude control, or housekeeping data sample is time stamped with a time word which is 48-bits long. Enumerating from the highest to lowest significant bit the word has a 2 bit server ID, a 14 bit Major Period counter and a 32 bit tick counter. The servers and clients maintain their own copy of the time word and increment it at 100~kHz using their local oscillators. The Major Period is incremented by the servers every 6 hours. Every $2^{14}$ ticks (6.1~Hz) the time servers broadcast a sync message consisting of the 34 most significant bits of the time word. On receipt of a valid sync message the client uses these 34 bits to rewrite its local time word and set the value of the 14 least significant bits to zero. Relative timing of the bolometer and attitude control systems is determined by the drift of the local oscillators over the period between sync messages, which is less than 8.2~$\mu$s.

In addition to the sync message on the time bus, the time servers send a sync message on the CANbus every 200~ms. This message includes the high 40 bits of the time word. When an ELMB on one of the boards receives the message it responds by sampling the data and writing it to the CANbus.

Absolute time information is obtained via GPS. The GPS sends a pulse every second, accurate to a few tens of nanoseconds, to each time server which then records its current time word to disk via CANbus. In the worst case of only a single GPS pulse near launch and for the anticipated 14 days of long duration flight, the stability of the clocks of the time servers should maintain an absolute accuracy of better than 3.6~ms. 

\section{Status and Outlook}
On June 11, 2009 EBEX was launched on an engineering test flight from NASA's Columbia Scientific Ballooning Facility in Fort Sumner, NM, spending 10 hours at an altitude above 35~km. EBEX was the first experiment to successfully operate arrays of TES bolometers read out by SQUID amplifiers in a space-like environment. SQUIDs and bolometers were tuned during the flight and maintained their stability. A HWP was continuously rotating on an SMB throughout the flight. The cryostat temperatures, as monitored by the housekeeping boards over the CANbus, were stable throughout the data taking portion of the flight. There was no evidence of cosmic ray hits in the housekeeping data, and the boards proved to be robust in a balloon environment. Preparations are underway for an Antarctic flight.

\section{Acknowledgments}
EBEX is a NASA supported mission through grants number NNX08AG40G and NNX07AP36H. We thank Columbia Scientific Balloon Facility for their enthusiastic support of EBEX. We also acknowledge support from CNRS, Minnesota Super computing Institute and the Science and Technology Facilities Council. This research used resources of the National Energy Research Scientific Computing Center, which is supported by the office of Science of the U.S. Department of Energy under contract No. DE-AC02-05CH11231. The McGill authors acknowledge funding from the Canadian Space Agency, Natural Sciences and Engineering Research Council, Canadian Institute for Advanced Research, Canadian Foundation for Innovation and Canada Research Chairs program.

\bibliographystyle{ws-procs975x65}
\bibliography{mybib}

\begin{thebibliography}{10}
\bibitem{cardiff}
Desiged and manufactured by our collaborators at Cardiff University
\end{thebibliography}

\end{document}